\documentclass{article}
\usepackage{waspaa23,amsmath,graphicx,url,times}
\usepackage{amssymb,amsfonts,mathrsfs}
\usepackage{algorithm,algorithmic}
\usepackage{booktabs}
\usepackage{multirow}
\usepackage{color}
\usepackage{caption} 
\usepackage{etoolbox,siunitx}
\usepackage{color}
\usepackage{hyperref}
\usepackage{threeparttable}

 \renewenvironment{thebibliography}[1]{
 \begin{oldthebibliography}{#1}
 \setlength{\itemsep}{0.0pt}
 \setlength{\parskip}{0.1em}
 }
 {
 \end{oldthebibliography}
 }

\usepackage{setspace}

\renewenvironment{thebibliography}[1]{
  \begin{oldthebibliography}{#1}
}
{
  \end{oldthebibliography}
}
\usepackage[
backend=biber,
style=ieee,
citestyle=numeric-comp,
maxbibnames=2,
maxcitenames=3,
doi=false,isbn=false,url=false,eprint=false
]{biblatex}
\defbibheading{bibliography}[\refname]{}

\title{Exploring the integration of speech separation and recognition with self-supervised learning representation}

\name{{\shortstack[c]{
      Yoshiki Masuyama,$^{1\star}$
      \thanks{$^\star$Equal contribution.}
      Xuankai Chang,$^{2\star}$
      Wangyou Zhang,$^{3}$
      Samuele Cornell,$^{4}$ \\
      Zhong-Qiu Wang,$^{2}$
      Nobutaka Ono,$^{1}$
      Yanmin Qian,$^{3}$
      Shinji Watanabe,$^{2}$}
      }}
\address{$^1$Tokyo Metropolitan University, Japan \,\,
         $^2$Carnegie Mellon University, USA \\
         $^3$Shanghai Jiao Tong University, China \,\,
         $^4$Università Politecnica delle Marche, Italy
}

\begin{document}

\ninept
\maketitle

\begin{sloppy}

\begin{abstract}
Neural speech separation has made remarkable progress and its integration with automatic speech recognition (ASR) is an important direction towards realizing multi-speaker ASR.
This work provides an insightful investigation of speech separation in reverberant and noisy-reverberant scenarios as an ASR front-end.
In detail, we explore multi-channel separation methods, mask-based beamforming and complex spectral mapping, as well as the best features to use in the ASR back-end model.
We employ the recent self-supervised learning representation (SSLR) as a feature and improve the recognition performance from the case with filterbank features.
To further improve multi-speaker recognition performance, we present a carefully designed training strategy for integrating speech separation and recognition with SSLR.
The proposed integration using TF-GridNet-based complex spectral mapping and WavLM-based SSLR achieves a $2.5\%$ word error rate in reverberant WHAMR! test set, significantly outperforming an existing mask-based MVDR beamforming and filterbank integration ($28.9\%$).
\end{abstract}


\section{Introduction}
\label{sec:intro}

Speech separation and enhancement (SSE) is a crucial front-end for various applications such as speaker diarization, automatic speech recognition (ASR), and spoken language understanding~\cite{Raj2021,li2017acoustic,lu2022espnet}.
The speech separation field has been revolutionized by the invention of deep clustering~\cite{Hershey2016} and permutation invariant training (PIT)~\cite{yu2017permutation}, which allow us to train deep neural networks (DNNs) for speech separation in a supervised manner.
Previous speech separation methods based on time-frequency (T-F) masking \cite{Hershey2016,Wang2018a,yu2017permutation,Wang2018} used a DNN to estimate the T-F mask for each speaker from the short-time Fourier transform (STFT) of the observed mixture.
Meanwhile, time-domain methods~\cite{Luo2019,Luo2020,Subakan2021} have demonstrated promising results by directly processing time-domain signals in an end-to-end (E2E) manner.
Recently, fully complex STFT-domain methods have been proven to be extremely effective~\cite{Yang2022,Tan2022a,Wang2022c}.
In particular, TF-GridNet \cite{Wang2022c} has achieved state-of-the-art (SotA) performance on several SSE benchmarks~\cite{Hershey2016,Wang2018a,Maciejewski2020}, including both monaural and multi-channel cases. 
Despite these impressive recent improvements in separation performance, it is still unclear how and when they can also lead to better ASR performance.

Most conventional SSE models are trained to minimize signal-level differences between the separated and target speech~\cite{Luo2019, Luo2020}.
This could lead to mismatches with respect to the subsequent ASR task.
To address this issue, several attempts~\cite{seltzer2004likelihood,li2016neural,heymann2017beamnet,Ochiai2017,minhua2019frequency,Chang2019,Zhang2021,von2020multi} have been made by integrating SSE and ASR models with joint optimization.
For robust ASR, a neural beamformer and a joint connectionist temporal classification
(CTC)/attention-based encoder-decoder were integrated and optimized with the ASR objectives~\cite{Ochiai2017}.
This integration was extended to multi-speaker settings including MIMO-Speech~\cite{Chang2019}.
It aims to directly improve the performance of multi-speaker ASR while preserving the modularity of the entire system, as opposed to a fully E2E black-box approach~\cite{seki2018end,Kanda2020,sklyar2021streaming}.
The intermediate separated speech achieves a good separation quality \cite{Chang2019}, although any signal-level criteria are not used for training.

Self-supervised learning (SSL) models such as Wav2Vec 2.0~\cite{Baevski2020}, HuBERT \cite{Hsu2021}, and WavLM~\cite{Chen2021} have shown considerable potential in a wide range of speech processing tasks~\cite{Yang2021,Tsai2022}.
Recently, IRIS~\cite{Chang2022} demonstrated impressive results with an E2E model that integrates monaural speech enhancement, WavLM, and ASR models.
MultiIRIS~\cite{Masuyama2023} expanded IRIS to perform multi-channel speech enhancement and demonstrated the effectiveness of the joint training under noisy and reverberant conditions.

\begin{figure}[t!]
\centering
\includegraphics[width=0.99\columnwidth]{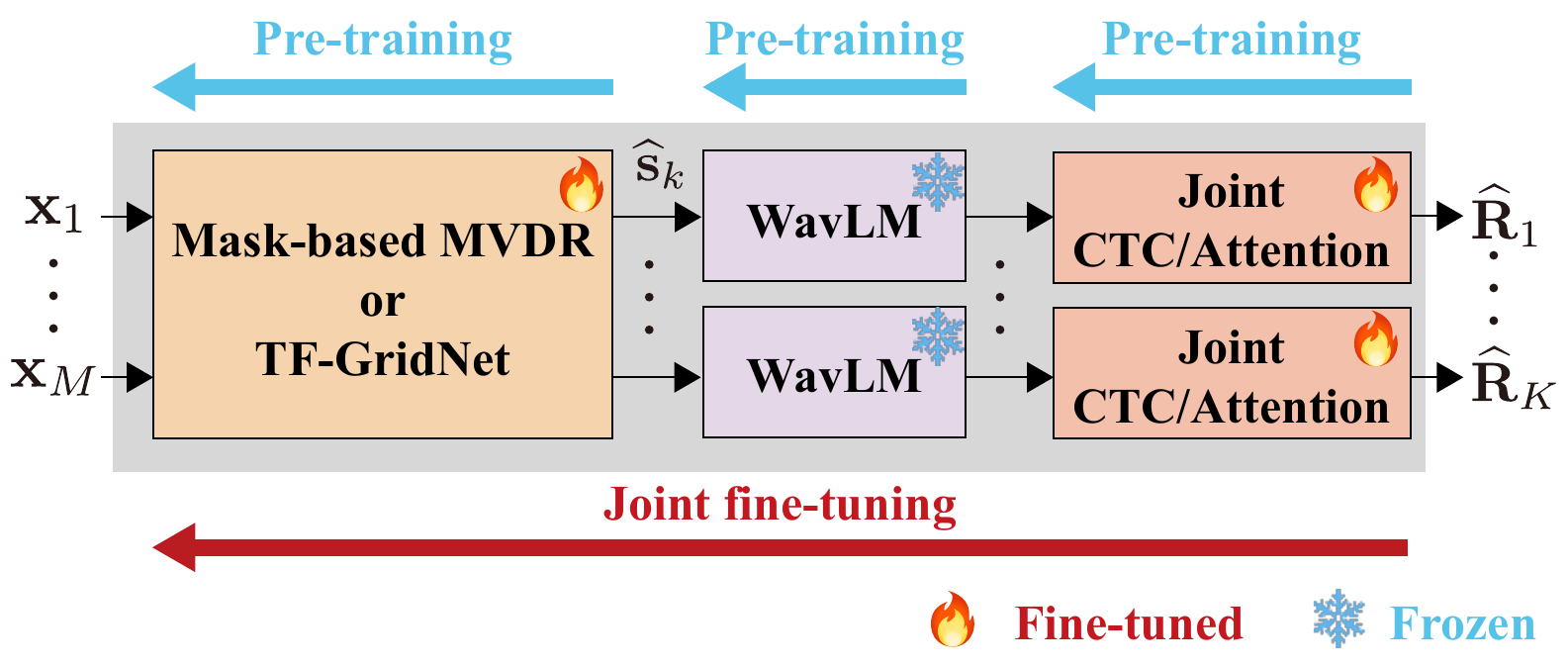}
\vspace{-4pt}
\caption{Overview of our E2E integration.
We pre-train speech separation, SSLR, and ASR models separately, and fine-tune the speech separation and ASR models jointly while freezing WavLM.}
\label{fig:overview}
\vspace{-6pt}
\end{figure}

Building upon MultiIRIS, this paper investigates MIMO-IRIS: an E2E integration of speech separation, SSLR extraction, and ASR for multi-channel multi-speaker overlapping scenarios.
We explore the combination of SSLR-based ASR models \cite{chang2021exploration} with TF-GridNet \cite{Wang2022c} as well as well-established beamforming techniques as illustrated in Fig.~\ref{fig:overview}.
We perform an extensive experimental validation on the spatialized WSJ0-2mix~\cite{Wang2018a} and WHAMR!~\cite{Maciejewski2020} datasets, assessing both separation and ASR performance.
Interestingly, our experiments show that the correlation between speech separation and ASR performance is not precisely positive.
The separation performance after fine-tuning degraded the separation performance while the word error rate (WER) decreases.
This is especially true for TF-GridNet-based complex spectral mapping, while mask-based beamforming~\cite{heymann2016neural,erdogan2016improved} results in less degradation.
Despite this, our best MIMO-IRIS model after joint training achieves SotA ASR performance on the WHAMR! dataset with a WER of $2.5\%$, comparable to SotA results on clean single-speaker WSJ evaluation sets \cite{chang2021exploration}.

\section{End-to-end Multi-Channel Multi-Speaker ASR with Speech Separation and SSLR}

Given an $L$-sample, $M$-channel mixture signal $\mathbf{X} = (\mathbf{x}_m)_{m=1}^M \in \mathbb{R}^{M \times L}$ consisting of $K$ speakers and noises $\mathbf{N} = (\mathbf{n}_m)_{m=1}^M$, we formulate the mixing process as follows:
\begin{equation}
    \mathbf{x}_m = \sum_{k=1}^{K} \mathbf{s}_{k, m} + \mathbf{n}_m,
\end{equation}
where $\mathbf{s}_{k,m} \in \mathbb{R}^{L}$ is the source image of speaker $k$ at microphone $m$.
The transcription sequence for speaker $k$ is denoted as $\mathbf{R}_k$.
This section describes each part of the proposed E2E system, depicted in Fig.~\ref{fig:overview}, including speech separation, SSLR extraction, and ASR.

\subsection{Speech Separation}
\label{ssec:separation}

The goal of speech separation is to estimate each speaker's signal $\mathbf{\hat{s}}_{k, r}$ at a reference microphone $r \in \{1, \dots, M\}$ from the mixture $\mathbf{X}$, which can be written as:
\begin{align}
\{\widehat{\mathbf{s}}_{1}, \dots, \widehat{\mathbf{s}}_{K}\} = \texttt{SS}(\mathbf{X}).
\label{eq:ss}
\end{align}
Depending on the number of input microphones, the task can be divided into monaural and multi-channel speech separation.

\vspace{-5pt}
\subsubsection{Monaural speech separation}
\label{sssec:monaural_ss}

While our main focus is on multi-channel speech separation, we briefly explain monaural speech separation as TF-GridNet was originally proposed for the monaural case.
In monaural speech separation, masking and mapping are two popular approaches  \cite{Wang2018}.
Both can be performed in the complex T-F domain or in the time domain.

In masking-based approaches, a DNN is trained to estimate a mask for each speaker, and the mask is point-wisely applied to the encoded representation of the mixture $\mathbf{X}$:
\begin{align}
    \mathbf{Z} &= \texttt{SSEnc}(\mathbf{X}), \\
    \{\widehat{\mathbf{G}}_{1}, \ldots, \widehat{\mathbf{G}}_{K}\} &= \texttt{MaskEstimationNet}(\mathbf{Z}), \label{eq:mask} \\
    \widehat{\mathbf{S}}_k &= \widehat{\mathbf{G}}_{k} \odot \mathbf{Z}, \label{eq:point_wise_mul} \\
    \widehat{\mathbf{s}}_{k} &= \texttt{SSDec}(\widehat{\mathbf{S}}_k),
    \label{eq:istft}
\end{align}
where $\widehat{\mathbf{G}}_{k}$ denotes the estimated mask for speaker $k$, and $\odot$ denotes the Hadamard product.
In T-F masking, $\texttt{SSEnc}$ and $\texttt{SSDec}$ are STFT and inverse STFT, respectively.
Meanwhile, they are usually trainable one-dimensional convolutional layers and deconvolutional layers in the time-domain methods.

In mapping-based approaches, a DNN is trained to directly predict the encoded representation of each speaker.
In detail, \eqref{eq:mask} and \eqref{eq:point_wise_mul} are replaced by
\begin{align}\label{eq:map}
    \{\widehat{\mathbf{S}}_1, \ldots, \widehat{\mathbf{S}}_K\} &= \texttt{MappingNet}(\mathbf{Z}).
\end{align}
The mapping-based approaches in the T-F domain, or complex spectral mapping, have gained increasing attention due to the appearance of powerful DNN architecture called TF-GridNet~\cite{Wang2022c}.
TF-GridNet predicts the real and imaginary components of each speaker from those of the observed mixture.
It has outperformed the best time-domain masking-based methods~\cite{Subakan2021}.
Furthermore, it has been successfully adapted to multi-channel speech separation.

\vspace{-5pt}
\subsubsection{Multi-channel speech separation}
\label{sssec:multich_ss}

Multi-channel speech separation takes advantage of spatial information afforded by multiple microphones and has been used in robust ASR~\cite{li2017acoustic,heymann2016neural,erdogan2016improved}.
For the purpose of robust ASR, two popular approaches have been developed multi-channel separation: using DNN estimates to derive a conventional beamformer and using DNN to directly estimate each speaker's signal.

In the first approach, the minimum variance distortionless response (MVDR) beamformer has been widely used due to its distortionless property and generalization capability~\cite{Gannot2017,heymann2016neural,erdogan2016improved,yosioka2018x}.
It incurs few processing artifacts by using the constrained time-invariant linear filters and is a preferable front-end of ASR backends~\cite{Chang2019, Zhang2021}. 
Neural mask-based beamforming estimates a T-F mask for each speaker $\widehat{\mathbf{G}}_k$ and computes a spatial covariance matrix as follows:
\begin{equation}
    \widehat{\mathbf{V}}_{k}[f] = \frac{1}{\sum_t \widehat{G}_k[t,f]} \sum_{t=1}^T \widehat{G}_k[t, f] \mathbf{z}[t,f] \mathbf{z}[t,f]^\mathsf{H},
\end{equation}
where $\mathbf{z}[t,f] = [Z_1[t,f], \ldots, Z_M[t,f]]^\mathsf{T}$, $Z_m[t,f]$ is the STFT coefficient of $\mathbf{x}_m$, $(\cdot)^\mathsf{T}$ denotes the transpose, and $(\cdot)^\mathsf{H}$ denotes the Hermitian transpose. 
An MVDR beamformer $\widehat{\mathbf{w}}_{k}[f]$ is given by
\begin{align}
    \hat{\mathbf{w}}_{k}[f] &= \frac{\widehat{\mathbf{V}}_{\backslash k}^{-1}[f] \widehat{\mathbf{V}}_{k}[f]}{\mathrm{trace}\Big(\widehat{\mathbf{V}}_{\backslash k}^{-1}[f] \widehat{\mathbf{V}}_{k}[f] \Big)} \mathbf{u},
\end{align}
where $\widehat{\mathbf{V}}_{\backslash k}[f]$ denotes the sum of the spatial covariance matrices of the noise and all the speakers except for speaker $k$, and $\mathbf{u} \in \mathbb{R}^{M}$ is a one-hot vector indicating the reference microphone.
The beamforming output is computed as:
\begin{align}
    \widehat{S}_k[t,f] &= \widehat{\mathbf{w}}_{k}^{\mathsf{H}}[f] \mathbf{z}[t,f], \label{eq:beamforming}
\end{align}
and converted to the time domain via inverse STFT as in \eqref{eq:istft}.

In the second approach, a DNN directly estimates the encoded representation of each speaker by replacing the input of \eqref{eq:map} to the concatenation of the encoded representation of microphone $m$.
Compared to the output of linear beamformers, the output of the second approach tends to have fewer non-target signals but more distortion on the target speech.
Although earlier studies suggested that linear beamformers would be preferable for robust ASR \cite{Chang2019, Zhang2021}, modern ASR back-ends and separation front-ends have become much more powerful nowadays.
Hence, we expect that modern back-ends could handle speech distortion in separated signals, and modern front-ends can produce much less distortion in separated signals.
We will compare their performance in our experiments, where TF-GridNet~\cite{Wang2022c} and the joint CTC/attention-based encoder-decoder~\cite{kim2017joint} are used for speech separation and ASR, respectively.

\subsection{SSLR Extraction and E2E-ASR}
\label{ssec:ssl_asr}

We extract SSLR from each separated signal $\widehat{\mathbf{s}}_k$ in \eqref{eq:ss} and pass it to E2E-ASR in the same way as in previous studies~\cite{Chang2022,Masuyama2023}:
\begin{align}
    \widehat{\mathbf{R}}_k &= \texttt{ASR}(\texttt{SSLR}(\widehat{\mathbf{s}}_k; \,\theta^{\text{ssl}}); \,\theta^{\text{asr}}),
    \label{eq:asr_ssl}
\end{align}
where $\theta^{\text{ssl}}$ and $\theta^{\text{asr}}$ represent the parameters of the SSLR extractor $\texttt{SSLR}(\cdot)$ and ASR model $\texttt{ASR}(\cdot)$, respectively.
Specifically, WavLM~\cite{Chen2021} is used to extract robust SSLR by applying the weighted sum of all transformer encoder embeddings.
The weights are optimized with the following ASR model.
E2E-ASR is based on the joint CTC/attention-based encoder-decoder framework~\cite{kim2017joint}.

\subsection{MIMO-IRIS: Integration of Separation, SSLR and ASR}
\label{ssec:mimo_iris}

To recognize multi-speaker speech, one can directly send the outputs of the speech separation model to a pre-trained ASR model.
This solution is, however, not optimal because ASR models are typically trained with single-speaker speech, while the separated speech usually contain residual interference.
Following IRIS~\cite{Chang2022} and MultiIRIS~\cite{Masuyama2023}, we integrate the speech separation model, SSLR extractor, and E2E-ASR model into a single model as shown in Fig.~\ref{fig:overview}.
The speech separation model can generate multiple streams, one for each speaker, and the ASR model is shared among all separated streams along with the SSLR extractor.
During the training, PIT is applied to the CTC loss in the ASR model to determine the optimal permutation.
The following attention-based decoder uses this permutation to select the corresponding reference transcript for each input stream in the teacher-forcing training.
Our E2E model can be extended from~\eqref{eq:asr_ssl} as:
\begin{align}
    & \{\hat{\mathbf{R}}_1, \dots, \hat{\mathbf{R}}_K\} = \texttt{ASR}(\texttt{SSLR}(\texttt{SS}(\mathbf{X};\theta^{\texttt{ss}}); \,\theta^{\texttt{ssl}}); \, \theta^{\texttt{asr}}),
    \label{eq:asr_ssl_ss}
\end{align}
where $\theta^{\texttt{ss}}$ represents the parameters of the speech separation model, as discussed in Section~\ref{ssec:separation}. The loss function of the ASR task is the same as in MIMO-Speech~\cite{Chang2019}. We omit the details here.

The E2E model could be trained from scratch with multi-task learning, including speech separation and ASR objectives.
Such training, however, requires intensive computation.
In addition, previous studies on the integration of speech enhancement, SSLR extraction, and E2E ASR reported that the integrated model resulted in sub-optimal performance when trained from scratch~\cite{Chang2022,Masuyama2023}.
We thus propose a two-stage approach.
First, the speech separation model is pre-trained on commonly-used speech separation datasets, e.g., spatialized WSJ0-2mix \cite{Hershey2016,Wang2018} and WHAMR! \cite{Maciejewski2020}.
Second, the ASR model is pre-trained on monaural clean speech datasets, e.g., the WSJ corpus.
Finally, the entire integrated model is fine-tuned with the ASR objective, as shown in Fig.~\ref{fig:overview}.
Following previous studies, we freeze the WavLM, which is pre-trained on a large amount of external data.
This strategy is efficient and requires only a few optimization epochs to achieve excellent performance in speech enhancement \cite{Chang2022,Masuyama2023}.

\section{Experiments}
\label{sec:exp}

We validate the effectiveness of our integration on two-speaker mixtures under anechoic/reverberant and clean/noisy conditions.
Our experiments were conducted using the ESPnet-SE++ toolkit%
\footnote{
Our source codes and configurations will be available through ESPnet:
\url{https://github.com/espnet/espnet}.
}%
\cite{lu2022espnet}.

\subsection{Datasets}

We evaluated our systems on the spatialized WSJ0-2mix~\cite{Wang2018a} and WHAMR!~\cite{Maciejewski2020} datasets, both of which support anechoic and reverberant two-speaker mixture simulations.
The training, validation, and test sets of both datasets contain 20,000, 5,000, and 3,000 mixtures, respectively. 
Room impulse responses were simulated and convolved with dry source signals from WSJ0-2mix~\cite{Hershey2016}.
The signal-to-distortion ratio (SDR)~\cite{vincent2006performance} with respect to the input mixture is 0.07\, dB in spatialized WSJ0-2mix.
WHAMR!~\cite{Maciejewski2020} is one of the most challenging datasets for speech separation, as it contains two-channel real-recorded environmental noise.
For WHAMR!, the SDR with respect to the input mixture is -4.61\, dB.
To leverage the pre-trained WavLM \cite{Chen2021}, which was trained on 16 kHz, we used the 16\,kHz version of both datasets in our experiments.
We combined both anechoic and reverberant conditions of the training and validation sets to form the new training and validation sets, respectively.

\subsection{Training Configurations}
\label{sec:spatialized-wsj0-2mix}

The ASR model ($\texttt{ASR}(\cdot)$ in~\eqref{eq:asr_ssl} and \eqref{eq:asr_ssl_ss}) consists of a Conformer-based encoder of 12 layers and a Transformer-based decoder of 6 layers by following a previous study \cite{Masuyama2023}.
The encoder and decoder have 2,048 hidden units and 4 attention heads.
We reduced the dimensions of the speaker-wise SSLR from 1,024 to 80 by a fully-connected layer before feeding it to the ASR model.
The ASR model and the learnable weight for the WavLM embeddings were pre-trained on the clean WSJ corpus.
We used the Adam optimizer with a warm-up and the peak learning rate of $1.0 \times 10^{-3}$.
During inference, we also used a Transformer-based character-level language model.
On the clean single-speaker WSJ evaluation set, the ASR model achieved a WER of $1.3\%$.

As the speech separation model ($\texttt{SS}(\cdot)$ in~\eqref{eq:ss} and \eqref{eq:asr_ssl_ss}), our mask-based MVDR beamformer employed a 3-layer bidirectional long short-term memory of 512 units with a projection layer to estimate the T-F masks as in \cite{Chang2019,Zhang2022}.
STFT was implemented with the Hann window of 512 samples with a 128-sample shift.
The mask estimation network was optimized with the convolutive transfer function invariant signal-to-distortion ratio (CI-SDR) loss~\cite{Boeddeker2021} on beamforming outputs.
Meanwhile, TF-GridNet consists of 6 blocks, where the TF-unit embedding dimension was $48$.
To reduce the computation, we increased the window shift size to 256 samples in STFT.
TF-GridNet was optimized with a sum of the $L_1$ loss on the waveform and on the STFT magnitude%
\footnote{
In our preliminary experiments, we also used the loss presented in \cite{Lu2022} to train the mask-based beamformer.
This resulted in worse WERs on the validation sets than using the CI-SDR loss~\cite{Boeddeker2021}.
}
following \cite{Lu2022}, where the weight for the waveform loss was 0.99.
Both mask estimation network and TF-GridNet were pre-trained with the Adam optimizer.
Then, the joint fine-tuning of the speech separation and ASR models was performed using the stochastic gradient descent method with a learning rate of $1.0 \times 10^{-3}$ and momentum of $0.9$.
We used the \textit{max} condition of the spatialized WSJ0-2mix and WHAMR! datasets, mixtures of the non-trimmed utterances, in the joint fine-tuning.

\subsection{Results on Clean Multi-channel Speech Separation}

Table~\ref{tab:clean_multi} presents the results on the spatialized WSJ0-2mix dataset.
First, we show the results of the cascaded monaural TF-GridNet and ASR performance, an SDR of 19.4 dB and a WER of 4.8\%.
It outperformed an existing cascaded system with a time-domain masking-based method~\cite{Neumann2020}.
We then show the results in multi-channel cases, where the speech separation models were fine-tuned with the ASR objective.
The TF-GridNet model consistently outperformed the MVDR beamformer not only in terms of SDRs but also in terms of WERs.
This result demonstrates that the unconstrained complex spectral mapping is advantageous as an ASR front-end when using modern speech separation models.
Furthermore, even the monaural TF-GridNet is more effective than the MVDR beamformer without joint fine-tuning.

To clarify the effectiveness of WavLM as a robust SSLR extractor, we evaluated the ASR model using filterbank features without joint fine-tuning.
According to the bottom row of Table~\ref{tab:clean_multi}, its WER was degraded to $28.2\%$ from $2.4\%$ with WavLM in the reverberant condition.
This result confirms the importance of the robust SSLR even with the powerful complex spectral mapping.
In the weighted sum for extracting SSLR, the weight concentrated on the last layer, around $0.83$, similar to previous studies~\cite{Chang2022,Masuyama2023}.

As an interesting finding, joint fine-tuning further reduced the WERs in both anechoic and reverberant conditions while degrading the separation performance.
This degradation was less severe for the MVDR beamforming as its output is constrained to be distortion-less.
Meanwhile, TF-GridNet-based unconstrained complex spectral mapping faced severe performance degradation, despite the better WER. 
In the anechoic case, the multi-channel TF-GridNet can achieve an SDR of 26.43 dB and a WER of 3.2\% without fine-tuning.
However, the separation performance dropped to 15.28 dB after joint fine-tuning.
In detail, we observed buzzy artifacts in the intermediate separated signals%
\footnote{
Examples of spectrograms and audio signals are available online: \url{https://yoshikimas.github.io/mimo-iris}.
}.

\begin{table}[t]
    \centering
    \caption{Separation and WER results on single-channel WSJ0-2mix and spatialized WSJ0-2mix.
    }
    \scalebox{0.85}[0.85]{
    \begin{tabular}{l|ccc|c}
        \toprule
        & SDR [dB] & PESQ & STOI & WER (\%) \\
        \midrule
        \multicolumn{5}{c}{\emph{Monaural}} \\
        \midrule
        Time-domain$^\star$~\cite{Neumann2020} & 13.8 & - & - & 22.9 \\
        TF-GridNet$^\star$ & 19.40 & 3.41 & 0.976 & 4.8 \\
        \midrule
        \multicolumn{5}{c}{\emph{Anechoic eight-channel}} \\
        \midrule
        MVDR (\textbf{proposed}) & 12.83 & 3.86 & 0.987 & 2.1 \\
        - w/o fine-tuning & 14.53 & 3.90 & 0.989  & 7.8 \\
        TF-GridNet (\textbf{proposed}) & 15.28 & 3.14 & 0.983 & \textbf{1.7} \\
        - w/o fine-tuning & \multirow{2}{*}{\textbf{26.43}} & \multirow{2}{*}{\textbf{4.09}} & \multirow{2}{*}{\textbf{0.995}} & 3.2 \\
        - w/o WavLM &  &  &  & 6.3 \\
        \midrule
        \multicolumn{5}{c}{\emph{Reverberant eight-channel}} \\
        \midrule
        MVDR (\textbf{proposed}) & 4.56 & 2.76 & 0.859 & 3.6 \\ 
        - w/o fine-tuning & 5.11 & 2.76 & 0.864 & 30.5 \\
        TF-GridNet (\textbf{proposed}) & 12.32 & 3.17 & 0.956 & \textbf{1.8}  \\
        - w/o fine-tuning  & \multirow{2}{*}{\textbf{18.81}} & \multirow{2}{*}{\textbf{3.89}} & \multirow{2}{*}{\textbf{0.983}} & 2.4 \\
        - w/o WavLM &  &  &  & 28.2 \\
        \bottomrule
    \end{tabular}
    }
    \label{tab:clean_multi}
\begin{tablenotes}[center]\footnotesize
    \item[*] \!\!\!\!\! $^\star$ The monaural models were not jointly fine-tuned.
\end{tablenotes}
\vspace{-4pt}
\end{table}

\subsection{Results on Noisy Multi-channel Speech Separation}

In this section, we present our experimental results of the WHAMR! dataset, which are summarized in Table~\ref{tab:noisy}.
In the top panel, we report the performance of monaural TF-GridNet on both noisy anechoic and reverberant conditions.
As with the results on the spatialized WSJ0-2mix, the monaural TF-GridNet outperformed the mask-based MVDR beamformer integrated with weighted prediction error dereverberation~\cite{Zhang2020a}.
The difference is even more significant due to the limitation of the number of microphones and noisy/reverberant characteristics of the data.

The best model overall is again the multi-channel TF-GridNet, which reached the best signal-level metrics before fine-tuning.
After joint fine-tuning, the SDR decreased significantly, but the WER improved by over 400\% relative factor in the noisy/reverberant condition.
The performance is outstanding with WERs of $2.3\%$ and $2.5\%$ in anechoic and reverberant conditions, respectively, which are close to the performance achieved on the clean WSJ dataset.
We also fine-tuned the ASR model while freezing the separation model, and its results are in the second bottom row of Table~\ref{tab:noisy}.
While it outperformed the model without fine-tuning, its WER did not reach that of the joint fine-tuning model.
This result confirms the advantage of the joint fine-tuning of both front-end and back-end.
We emphasize that the ASR performance without fine-tuning still outperformed the previous MIMO-Speech~\cite{Zhang2020a} and the cascade combination of the time-domain speech separation and ASR models~\cite{Zhang2021a}.

\begin{table}[t]
    \centering
    \caption{Separation and WER results on WHAMR!.}
    \scalebox{0.85}[0.85]{
    \begin{tabular}{l|cc|cc}
        \toprule
         & \multicolumn{2}{c|}{Noisy/Anechoic} & \multicolumn{2}{c}{Noisy/Reverberant} \\
        \cmidrule{2-5}
        & SDR [dB]\!\! & \multicolumn{1}{c|}{WER (\%)} & SDR [dB]\!\! & WER (\%) \\
        \midrule
        \multicolumn{5}{c}{\emph{Monaural}} \\
        \midrule
        TF-GridNet$^\star$ & 9.27 & 14.5 & 9.07 & 18.3 \\
        \midrule
        \multicolumn{5}{c}{\emph{Two-channel}} \\
        \midrule
        MIMO-Speech~\cite{Zhang2022} & - & - & -2.27 & 28.9 \\
        Time-domain~\cite{Zhang2021a} & - & - & - &  20.9 \\
        MVDR (\textbf{proposed}) & -1.42 & 42.2 & -1.30 & 44.4 \\
        TF-GridNet (\textbf{proposed}) \!\!\!\! & 9.11 & \textbf{2.3} & 7.84 & \textbf{2.5} \\
        - ASR-only fine-tuning  \!\!\!\! & \multirow{2}{*}{\textbf{13.12}} & 4.4 & \multirow{2}{*}{\textbf{11.05}} & 6.5 \\
        - w/o fine-tuning & & 6.5 & & 10.5 \\
        \bottomrule
    \end{tabular}
    }
    \label{tab:noisy}
\begin{tablenotes}[flushleft]\footnotesize
    \item[*] $^\star$ The monaural TF-GridNet was not jointly fine-tuned.
\end{tablenotes}
\end{table}

\section{Conclusion}
\label{sec:conclusion}

In this paper, we investigated the integration of speech separation, SSLR extraction, and ASR with well-established beamforming techniques as well as the latest SotA techniques including TF-GridNet. 
Our experiments were perfromed under anechoic/reverberant and clean/noisy conditions using the spatialized WSJ0-2mix and WHAMR! datasets. 
In detail, we explored how both separation performance and WER are affected by joint fine-tuning. 
Our experimental results show that the purely DNN-based speech separation method, TF-GridNet-based complex spectral mapping, can considerably outperform the mask-based MVDR beamforming preferred as an ASR front-end. 
Joint fine-tuning degraded the separation performance while significantly improving the WER, which is inconsistent with the tendency reported in a speech enhancement paper~\cite{Masuyama2023}.
Our future work should focus on how this degradation can be prevented, e.g. by using continual learning strategies. 
Overall our best system, based on multi-channel TF-GridNet, WavLM, and E2E ASR, was able to reach performance on par with the one achieved on clean, single-speaker WSJ~\cite{chang2021exploration}.

\section{Acknowledgements}

Y. Masuyama was partially supported by JSPS KAKENHI Grant Numbers JP21J21371 and JST CREST Grant Number JPMJCR19A3.
X. Chang, Z.-Q. Wang, and W. Zhang used the Bridges2 system at PSC and Delta system at NCSA through allocation CIS210014 from the Advanced Cyberinfrastructure Coordination Ecosystem: Services \& Support (ACCESS) program.
S. Cornell was partially supported by Marche Region within the funded project ``Miracle'' POR MARCHE FESR 2014-2020.

\clearpage
\section{References}
{\footnotesize
\begingroup
\setstretch{0.75}
\setlength\bibitemsep{1.6pt}
\printbibliography
\endgroup
}

\end{sloppy}
\end{document}